\documentclass[preprint,showpacs,preprintnumbers,amsmath,amssymb]{revtex4}
\usepackage{dcolumn}
\usepackage{amssymb}
\usepackage{bm}
\usepackage{epsfig}
\usepackage{psfrag}
\usepackage{ulem}

\def\3dots{\:\raisebox{-0.5ex}{$\stackrel{\textstyle.}{:}$}\:}
\def\beq{\begin{equation}}
\def\eeq{\end{equation}}
\def\bea{\begin{eqnarray}}
\def\eea{\end{eqnarray}}

\begin{document}

\title{Formation of p-n junction in polymer electrolyte-top gated bilayer graphene transistor}

\author{Biswanath Chakraborty, Anindya Das}
\author{A. K. Sood}%
 \email{asood@physics.iisc.ernet.in}
\affiliation{Department of Physics, Indian
Institute of Science , Bangalore - 560012, India}

\begin{abstract}
We show simultaneous p and n type carrier injection  in bilayer graphene channel by varying the longitudinal bias across the channel and the top gate voltage. The top gate is applied electrochemically using solid polymer electrolyte and the gate capacitance is measured to be 1.5 $\mu F/cm^2$, a value about 125 times higher than the conventional SiO$_2$ back gate capacitance. Unlike the single layer graphene, the  drain-source current does not saturate on varying the drain-source bias voltage. The energy gap opened between the valence and conduction bands using top and back gate geometry is estimated.

\end{abstract}
\maketitle 
-

\section{Introduction}

In recent years, single and a few layer graphene are at the center stage of intense research to unravel physical properties of these new forms of carbon, with an eye on device applications. Ultra high mobilities and observation of ballistic transport make graphene based devices a potential alternative for silicon based devices \cite{Novoselov  04,Novoselov Nat 05,Rise Nov,Kim Suspended}. The zero energy gap between the valence and conduction band  in the energy spectrum of single layer graphene (SLG) makes it difficult to achieve a high on-off ratio in field effect transistor (FET). On the other hand, bilayer graphene (BLG) devices hold greater promise in terms of better device performance with larger on-off ratio. This is feasible because of the band gap opening due to the breaking of inversion symmetry of the two layers in an otherwise gapless energy band structure. Moreover, this gap between the conduction and valence band can be tuned by means of external gate electric field, as demonstrated both  theoretically \cite{GeimBiased,McannPRB(R), McannPRL, CastroNetoReview,MauriCondMat} and experimentally \cite{OhtaScience,Morpugo,Atin,NatureIR}. In particular , Zhang et al. \cite{NatureIR} have reported a direct observation of the band gap in the bilayer by infrared absorption. They have used dual gate bilayer FET where 80 nm Al$_2$O$_3$ film was used as the top gate and 285 nm SiO$_2$ as the back gate.

Misewich et al.\cite{CNTScience} have shown that it is possible to get polarized infrared optical emission from a carbon nanotube FET by creating an effective forward biased p-n junction along the nanotube channel. This was achieved by applying the gate-drain voltage V$_{GD}$, comparable to the drain-source voltage V$_{DS}$, and maximum emission was obtained when V$_{GD} = \frac{V_{DS}}{2}$. Electrical measurements showed that the optical emissions results from the radiative recombination of electrons and holes near the junction. Similar experiments carried out by Meric et. al. \cite{KimSaturation} on single layer graphene (SLG) device have reported formation of p-n junction along the graphene channel. However, owing to the absence of the band gap in SLG, the recombination of electrons and holes at/near the junction does not yield any radiation. In comparison, bilayer graphene (BLG) device can serve as a prospective candidate is this regard since a band gap can be opened and controlled simultaneously by application of external electric field perpendicular to the layers. Thus, by creating a  p-n junction along the bilayer channel by proper biasing of the device, we may be able to fabricate novel source of terahertz radiation \cite{TeraMag, TeraJETP} based on the recombination of carriers. 

In this paper we show that the BLG device can change from unipolar state to ambipolar state  by varying the drain-source bias along the bilayer channel at different top gate voltages (V$_{T}$). This has been  possible because of very high gate capacitance of the top gate. We also report a quantitative analysis of the band gap opening in the bilayer graphene energy spectrum. In order to control the gap and the position of the Fermi level independently, we have used the top and back gate device geometry. Solid polymer electrolyte and 300nm thick SiO$_2$ were used as top gate and back gate materials, respectively.  This top gate arrangement allows us to shift the Ferrmi energy significantly by applying a very small top gate voltage ($\sim$ 1 V) because of its higher gate capacitance due to nanometer thick Debye layer\cite{AndyNature}. In addition, our top gate geometry is a very simple way as compared to depositing high $\kappa$ dielectric electric materials like HfO$_2$ and Al$_2$O$_3$, using atomic layer deposition technique. We determine the top gate capacitance  accurately by an application of top and back gates on a bilayer graphene device. It is shown that our device has very high gate capacitance $\sim 1.5 \mu F/ cm ^{2}$, which is nearly 125 times larger compared to the gate capacitance of 300 nm SiO$_2$ (12 $nF/ cm^{2}$). 

\section{Experiments and Results}

\begin{figure}[htbp]
\resizebox{130mm}{!}{
\includegraphics{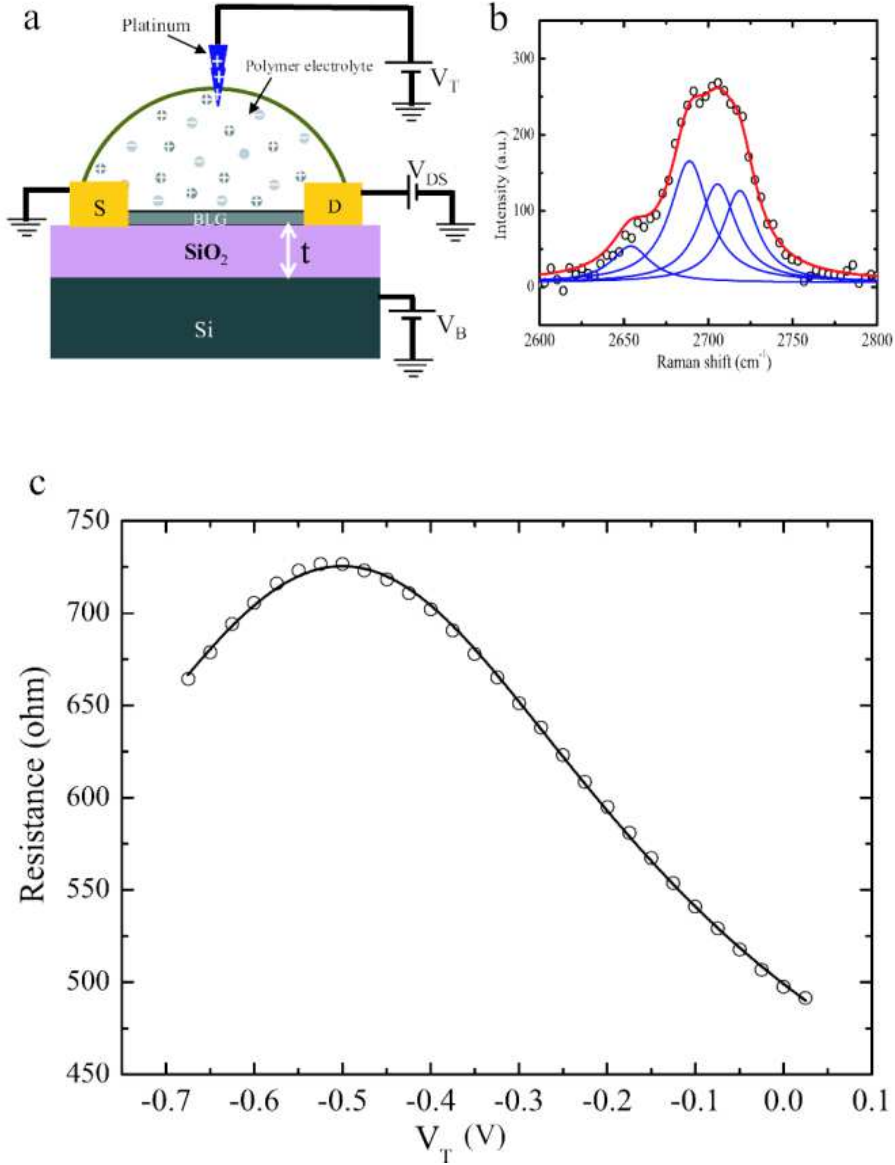}
}
\caption{(Color online) (a) Schematic of the experimental set up showing the top and back gate arrangement. (b) Raman spectrum of the bilayer graphene showing its characteristic 2D modes. (c) Resistance vs top gate voltage (V$_{T}$). Open circles are  experimental data and the solid line is fit to equation (\ref{fit}).}
\label{Fig1a}
\end{figure}

The device consists of bilayer graphene flake prepared from micromechanical cleavage of highly oriented pyrolytic graphite (HOPG) and deposited on Si/SiO$_{2}$ substrate with the oxide thickness (t) of 300 nm. The electrical contacts on the bilayer device were made by e-beam lithography, followed by  thermal evaporation of 30 nm of gold and subsequent liftoff in acetone. The source drain separation (L) is $\sim$500nm and the width (W) is $\sim$5$\mu$m. All the measurements were done at room temperature using two probe contacts. The drain-source and both the top and back gate biases were given from Keithley 2400 source meters. The top gate material solid polymer electrolyte consists of LiClO$_{4}$ and  polyethylene oxide (PEO) in the ratio 0.12:1 \cite{AndyNature, AndyPrb, AndyCnt}. The top gate voltage (V$_T$) was applied by inserting a platinum electrode within the polymer layer. Figure \ref{Fig1a}(a) shows a schematic of the experimental set up.  Raman spectroscopy using 514.5 nm excitation laser was used to identify the bilayer graphene (BLG) \cite{Ferrari fingerprint}. Raman spectrum of the bilayer graphene with the 2D mode resolved into four peaks is shown in figure \ref{Fig1a}(b). Figure \ref{Fig1a}(c) shows the ambipolar nature of I$_{DS}$ as a function of V$_{T}$ for V$_{DS}$ = 10 mV and the back gate voltage V$_B$ = 0 V. The Dirac point is at V$_{T}^{D}$ = -0.5 V, implying that our starting sample was electron doped. Mobility ($\mu$), the minimum carrier concentration ($\delta n$) and the total contact resistance (R$_C$)are extracted by fitting the data to an equation based on diffusive transport model,
\begin{eqnarray}
\label{fit}
&&R= R_C + R_{G} \nonumber \\
&&  = R_C + \frac{L}{W}\frac{1}{e\mu\sqrt{(\delta n)^2 + (n_{T})^2}}
\end{eqnarray} 
where R$_{G}$ is the resistance of the bilayer graphene, $n_{T}$ is the carrier density induced by the top gate and $\mu$, $\delta n$ and R$_C$ are  the fitting parameters and the contact resistance (R$_C$) is mainly at the source and drain contacts of the graphene and $\delta n$, arises mainly from charge puddles which are created due to charge impurities ($n_{imp}$) \cite{18 of KimSaturation, 19 of KimSaturation} lying between dielectric and the bilayer sample. In order to determine $n_{T}$, we estimate the value of the top gate capacitance C$_{T}$ as follows. 

\subsection{Determination of polymer gate capacitance C$_T$}

\begin{figure}[htbp]
\resizebox{0.7\textwidth}{!}{
\includegraphics{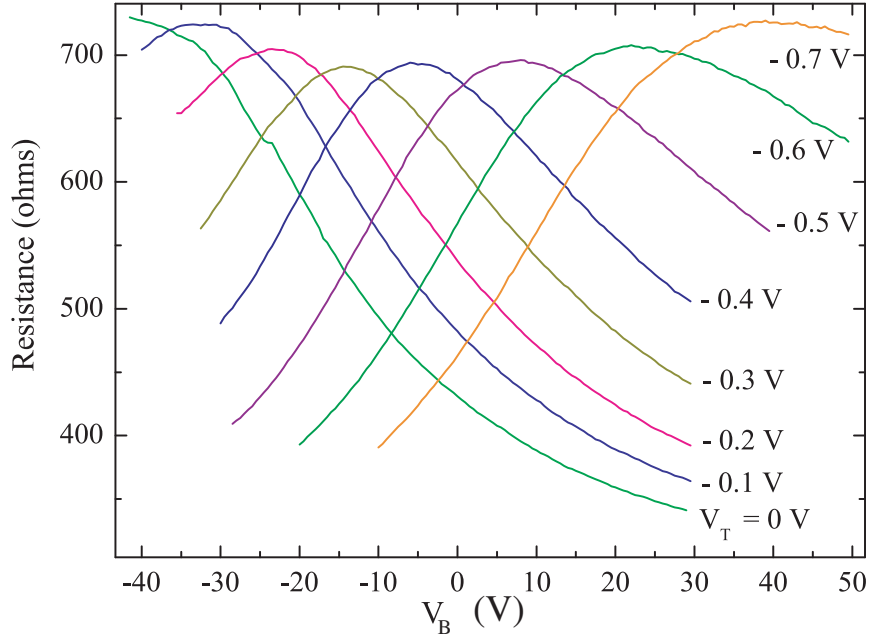}
}
\caption{(Color online) Resistance vs back gate voltage (V$_{B}$) for a fixed value of V$_{T}$.}
\label{Fig1}
\end{figure}

Figure \ref{Fig1} shows the resistance of the channel as a function of V$_{B}$ for a fixed V$_{T}$. For a given gate voltage V$_G$, 
\begin{equation}
\label{anygate}
V_G = \frac{E_F}{e} + \phi = \frac{E_F}{e} + \frac{ne}{C_G}
\end{equation}
where $\frac{E_F}{e}$ is determined by the quantum capacitance of the bilayer and potential $\phi$ from the geometrical capacitance of the gate\cite{AndyNature, AndyPrb}. In equation (\ref{anygate}), for top gate C$_G$=C$_T$, V$_G$=V$_T$, $n=n_T$ ; similarly for back gate C$_G$=C$_B$, V$_G$=V$_B$,  $n=n_B$. The gate-induced carrier density in a bilayer graphene is given by \cite{AndoBilayer, AndyPrb} 
\begin{equation}
 \label{n}
  n = \alpha[\gamma_1E_F + E_F^2], \hspace{3mm} E_F < \gamma_1
 \end{equation}
where $\gamma_1$ is the inter-layer hopping energy ($\sim$ 390 meV), $\alpha = \frac{1}{\pi({\hbar}{v_F})^2}$ and $v_F$ is the Fermi velocity = 10$^6 m/sec$ \cite{Rise Nov}. Therefore, for a given V$_{T}$, the induced carrier density in the bilayer sample  is 
\begin{equation}
\label{ntopgate}
n_{T} = \alpha\left(\gamma_1 A + A^2\right) = f(C_{T}, V_{T})
\end{equation}
where 
\begin{equation}
\label{A}
A = -\frac{1}{2}(\gamma_1+\frac{C_T}{\alpha e^2}) + \frac{1}{2}\sqrt{(\gamma_1+\frac{C_T}{\alpha e^2})^2 + \frac{4C_TV_T}{\alpha e}}
\end{equation}
Similar expressions hold good for the back gate geometry and when
resistance is maximum (R$_{m}$) at a back gate voltage $V_B^{max}$, $\vert{n_{T}}\vert = \vert{n_{B}}\vert$ to make the Fermi energy shift zero. Therefore, for a given V$_{T}$ 
\begin{equation}
\label{fn}
f(C_{T}, V_{T}) = f(C_{B}, V_{B}^{max})
\end{equation}
where C$_{T}$ is the only unknown parameter. Using the known C$_{B} = 12 nF/cm^{2}$ and $V_{B}^{max}$ for different V$_{T}$, the average value of C$_{T}$ is found to be $\sim1.5 \mu F/cm^{2}$. Note that, in this estimation the quantum capacitance is not assumed to be a constant as was done by Meric et. al.\cite{KimSaturation}. Knowing C$_{T}$, $n_{T}$ is thus determined as a function of V$_T$ from equation (\ref{ntopgate}) and (\ref{A}). The experimental curve of figure \ref{Fig1a}c is then fitted to equation (\ref{fit}) giving, $\mu$ = 730 cm$^2$/V.sec, $\delta n$ = 1.5 $\times 10^{12}/ cm^{2}$ and R$_C$ = 150 $\Omega$.

\subsection{Formation of p-n junction along the bilayer channel}

\begin{figure}[htbp]
\resizebox{0.7\textwidth}{!}{
\includegraphics{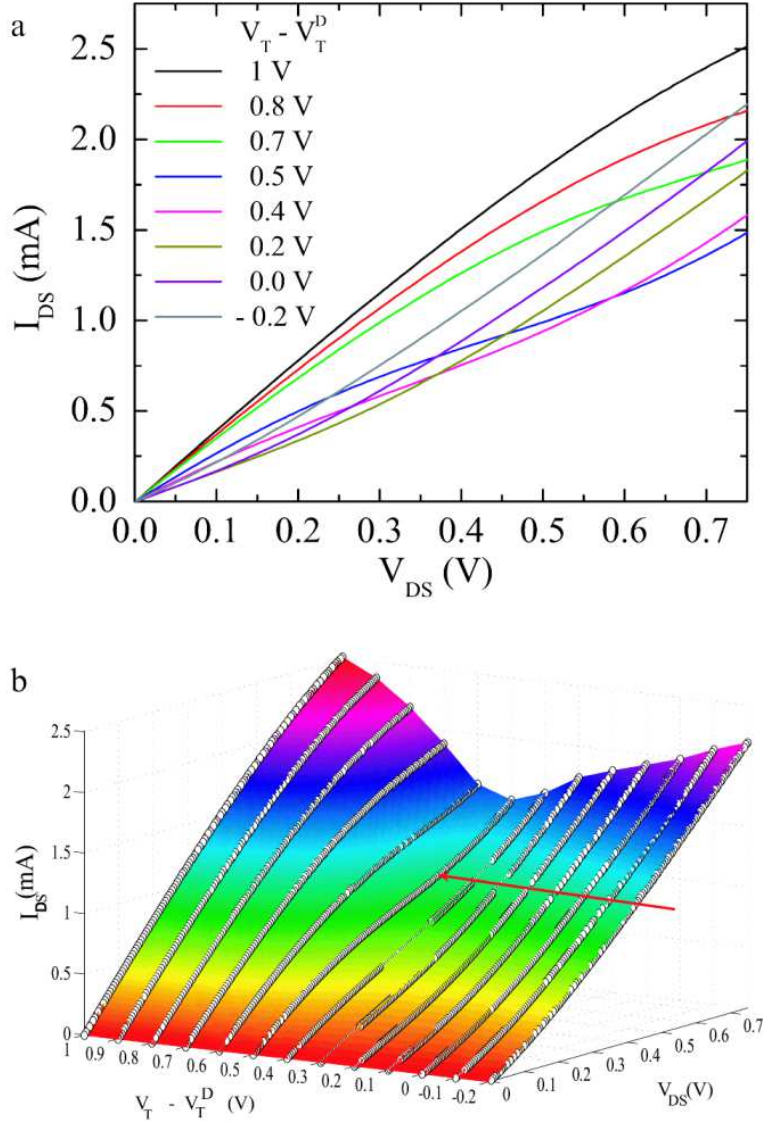}
}
\caption{(Color online) (a) I$_{DS}$ versus (V$_{DS}$) as a function of V$_T$ - V$_{T}^D$ for V$_B$ = 0V. (b) 3D plot showing the variation of I$_{DS}$ as function of both  $_{T}$ and V$_{DS}$. The point of inflection is shown by an arrow.}
\label{Fig3}
\end{figure}

We now discuss the characteristics of drain-source current (I$_{DS}$) as a function of drain-source bias (V$_{DS}$). Figure \ref{Fig3}(a) and b shows I$_{DS}$ vs V$_{DS}$ for different top gate voltages. The main noticeable feature is a clear non-linear dependence of I$_{DS}$ on V$_{DS}$, and a significant effect of the gate voltage on the shape of the I$_{DS}$ - V$_{DS}$ curves.  

\begin{figure}[htbp]
\resizebox{0.7\textwidth}{!}{
\includegraphics{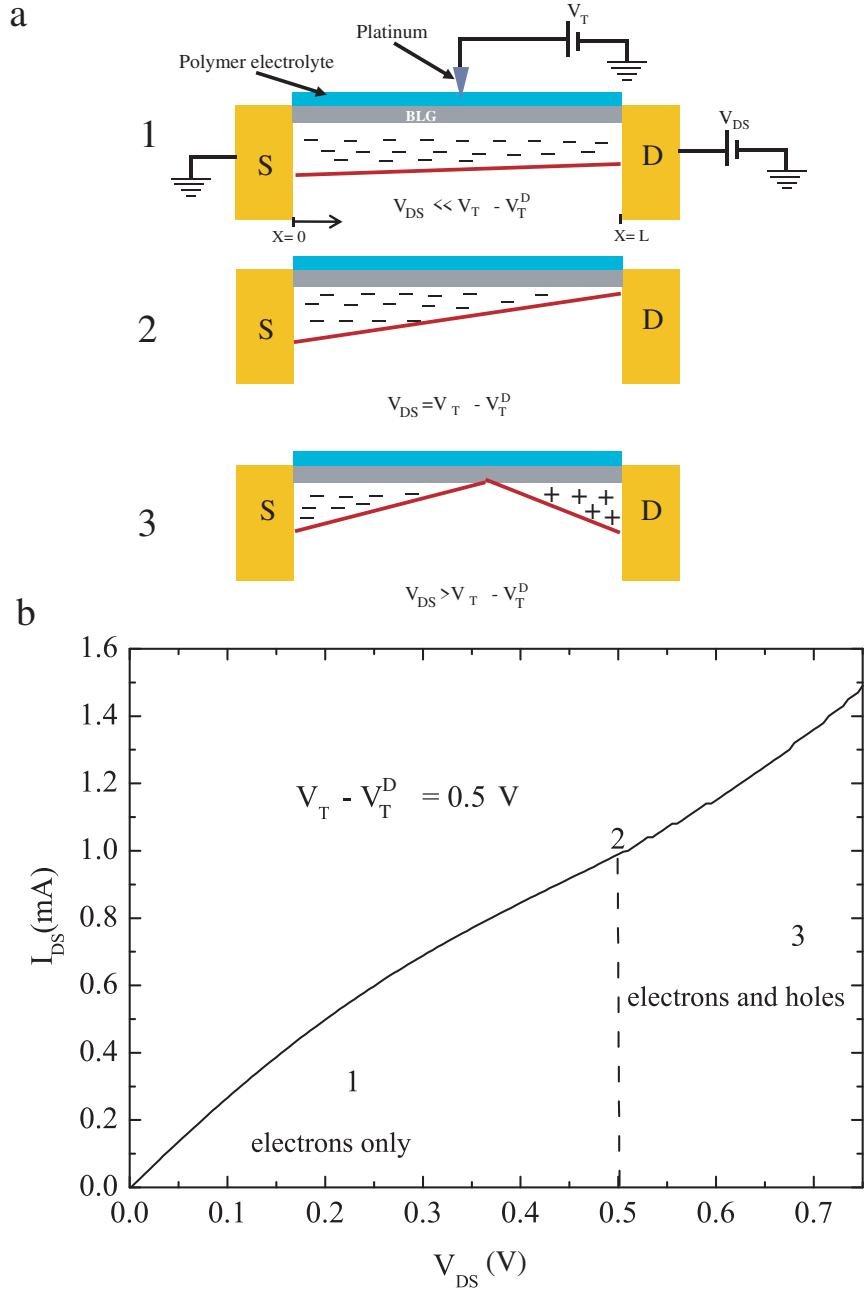}
}
\caption{(Color online) (a) Schematic of the spatial profile of the carrier concentration (n(x)) along the channel for a particular top gate voltage. Three cases are shown: Case 1: (V$_{DS}$ $<<$ V$_{T}$ - V$^{D}_{T}$) when the carriers are only electron; Case 2: V$_{DS}$ $=$ V$_{T}$ - V$^{D}_{T}$ when the pinch off starts appearing near the drain end and case 3: V$_{DS}$ $>$ V$_{T}$ - V$^{D}_{T}$ where the minimum carrier density point (pinch off) shifts towards the source end. (b) I$_{DS}$-V$_{DS}$ curve for V$_{T}$ - V$^{D}_{T}$ $=$ 0.5V pointing out the regions mentioned in Fig.\ref{Fig4}a. An inflection point appears at V$_{DS}$ $=$ V$_{T}$ - V$^{D}_{T}$ $=$ 0.5 V implying pinch off at drain end.}
\label{Fig4}
\end{figure}

To understand the non-linear I$_{DS}$ - V$_{DS}$ curve, we look at the schematic in figure \ref{Fig4}(a). For  V$_{DS}$ $\ll$ V$_{T}$, the bilayer channel will be electron doped for  (V$_{T}$ - V$_{T}^{D}$)$>$ 0 and hole doped for  (V$_{T}$ - V$_{T}^{D}$)$<$ 0 and the doping ($n(x)$) will be homogeneous along the channel length. However, this is not the case when V$_{DS}$ is comparable to V$_{T}$. In this case the voltage difference between the gate and the bilayer channel varies from (V$_{T}$ - V$_{T}^{D}$) at the source to V$_{DS}$ - (V$_{T}$ - V$_{T}^{D}$) at the drain electrode. Therefore, by varying drain-source voltage (V$_{DS}$) at fixed V$_{T}$, the doping concentration $n(x)$ changes from n type to p type along the channel length. In figure \ref{Fig4}(a), we have shown the carrier distribution $n(x)$, along the bilayer channel for three different cases of drain-source voltage (V$_{DS}$) at (V$_{T}$ - V$_{T}^{D}$) = 0.5 V.

Case 1: V$_{DS}$ $<$ (V$_{T}$ - V$_{T}^{D}$), and therefore the carriers $n(x)$ will  be electrons in the bilayer channel with $n(x = 0)$ $>$ $n(x$ = L). In this region, I$_{DS}$ increases sub linearly with V$_{DS}$ because the average carrier concentration [$<n(x)>$ = $\frac{1}{L}\displaystyle \int_{0}^{L}n(x)dx$] decreases with V$_{DS}$.

Case 2:  V$_{DS}$ = (V$_{T}$ - V$_{T}^{D}$), $n(x)$ will be zero at $x$ = L and therefore, the conduction channel gets pinched off near the drain end making the region devoid of charge carriers. At this point the slope of the I$_{DS}$ vs V$_{DS}$ curve undergo a change (figure \ref{Fig4}(b)). 

Case 3:  V$_{DS}$ $>$ (V$_{T}$ - V$_{T}^{D}$), the point in the channel where the pinch off occurs, moves deeper into the channel drifting towards the source electrode. As  V$_{DS}$ $>$ (V$_{T}$ - V$_{T}^{D}$), the gate is negatively biased with respect to drain near the drain region. As a result carrier concentration $n(x)$ will be holes near the drain region, as shown in the lower most panel of figure \ref{Fig4}(a). When V$_{DS}$ is increased beyond pinch off, the current increases due to enhanced p channel conduction.

Thus the bilayer channel is now viewed as comprising of distinct electron and hole regions, with the electron doped region shrinking as the applied drain-source bias is increased. The potential drop across the electron channel remains fixed at V$_{T}$ - V$_{T}^{D}$ while the drop across the hole region (from drain end to pinch off) increases as V$_{DS}$ - (V$_{T}$ - V$_{T}^{D}$), with the applied drain-source bias. As the device is operated beyond the pinch off, the dominant current carrier changes from electron to holes and the current increases as a function of V$_{DS}$. It appears as if a forwrad biased p-n junction is formed at the pinch off region, with p-region towards the drain end and n-region to wards the source region. 

A noticeable feature is the absence of current saturation in all the I$_{DS}$ vs V$_{DS}$ curves as compared to the monolayer case reported recently \cite{KimSaturation}. In order to address if the small length  of the channel in our experiment ($\sim$500nm) is responsible for non-saturation of current \cite{KimSaturation, CNTDrainScaling}, we carried out experiments on a device of drain-source separation $\sim$ 2.5$\mu$m and width $\sim$ 1.5 $\mu$m. This also did not show any signature of current saturation. 

\begin{figure}[htbp]
\resizebox{0.7\textwidth}{!}{
\includegraphics{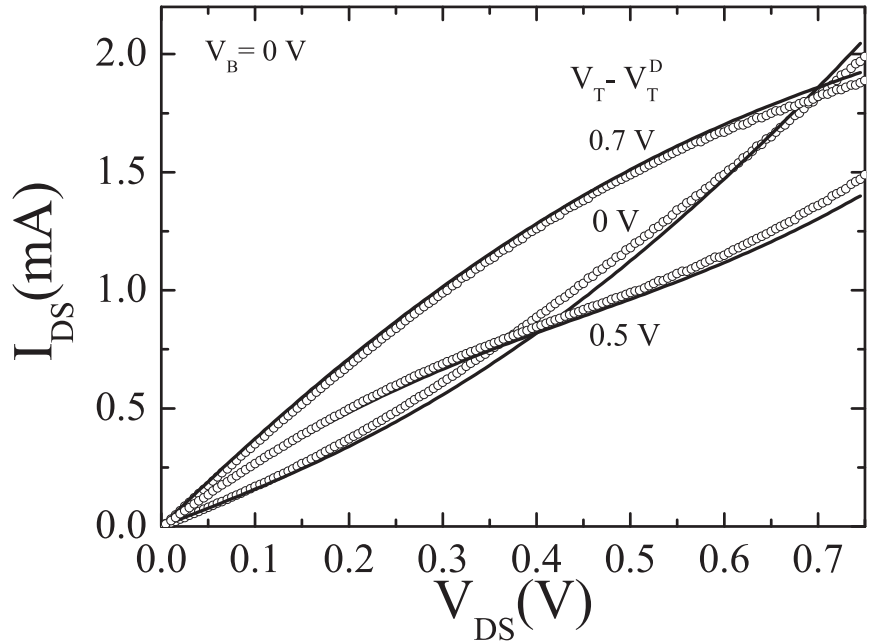}
}
\caption{(Color online) Comparison between experiment(open circles) and  theory(solid lines) is done for three gate voltages.}
\label{Fig5}
\end{figure}

In order to analyze the I$_{DS}$ - V$_{DS}$ curves, we model the drain ccurrent  as 

\begin {equation}
\label{1stIds}
I_{DS} = \frac{W}{L}\displaystyle \int_{0}^{L}ev_{D}(x)\sqrt{(\delta n)^2 + (n_{T}(x))^2}dx
\end{equation}

where $v$$_{D}$($x$) is drift velocity. 
The drift velocity can be written as 
\begin{equation} 
\label{driftvelocity}
v_D(x)= \mu E(x) = \mu \frac{dV(x)}{dx}
\end{equation}
where $E$ is the longitudinal electric field due to drain-source bias, V($x$) is the potential drop at point $x$ in the channel and $\mu$ is the mobility. 
Therefore,
\begin {equation}
\label{final Ids}
I_{DS} = \frac{W}{L}e\mu\displaystyle \int_{I_{DS}\frac{R_C}{2}}^{V_{DS}-I_{DS}\frac{R_C}{2}}\sqrt{(\delta n)^2 + (n_{T})^2}dV 
\end{equation}
where the upper and lower limits signifies the voltage drop at the drain and source ends, respectively. The drop due to the contact resistance at both the ends has been assumed to be  equal. Using the extracted values of  the mobility ($\mu$), the contact resistance (R$_{C}$), $\delta$n and $n_{T}$, equation (\ref{final Ids}) is solved numerically. Figure \ref{Fig5} shows experimental data (open circles) for three top gate voltages along with the theoretical curves (solid lines). It can be noted that agreement between the experiment and the calculation is quite good.

\subsection{Opening of energy gap in bilayer graphene and its estimation}

Coming back to figure \ref{Fig1}, it shows resistance as a function of the back gate voltage  for a fixed top gate voltage. For V$_T$ = 0 V, the maximum value of resistance (R$_{m}$) at a negative back gate voltage (V$_{B} = $V$_{B}^*$ $\sim -40 V$) clearly establishes that the bilayer graphene channel is intrinsically n-doped. Figure \ref{Fig1} shows that the value of R$_{m}$ is minimum at $V_{T}$ = -0.4 V and increases almost symmetrically on changing V$_T$.  It can also be seen, that for a particular top gate voltage, the drain-source current is not symmetric for  electron and hole doping.

\begin{figure}[htbp]
\resizebox{110mm}{!}{
\includegraphics{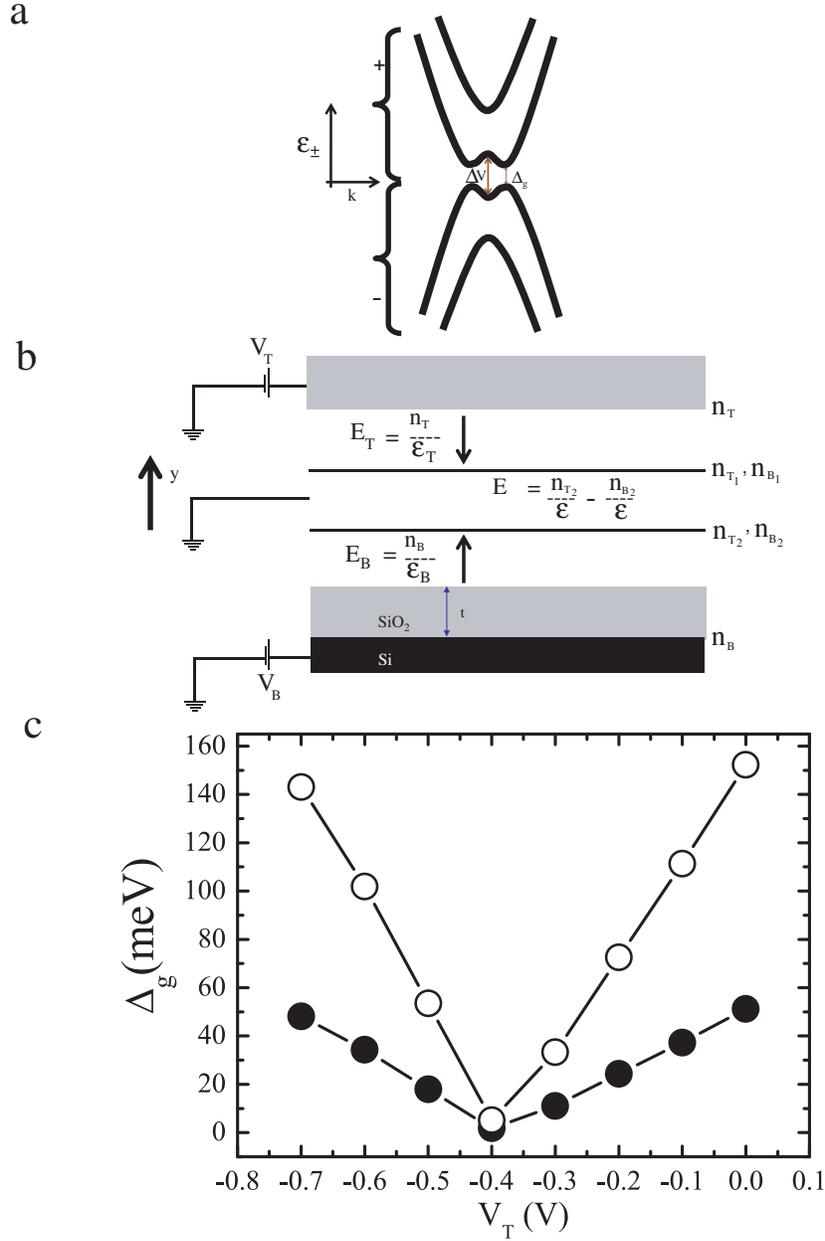}
}
\caption{(Color online) (a) Band structure of the biased bilayer graphene. (b)Schematic of the electric fields between the carbon layers  The arrows show the direction of the electric fields. (c) Energy gap $\Delta_g$ as a function V$_{T}$. The open (filled) circles shows the vales of the gap without (with) screening.}
\label{FigGap}
\end{figure}

The change in the maximum value of  resistance (R$_{m}$) at different top gate voltages arises due to the opening of an energy gap between the conduction and valence bands of the bilayer graphene \cite{GeimBiased, OhtaScience, Morpugo,  Atin, NatureIR}. It has been shown both theoretically\cite{McannPRB(R), McannPRL, CastroNetoReview}  and experimentally \cite{GeimBiased, OhtaScience, Morpugo,  Atin, NatureIR} that a difference between the on-site energy in the layers leads to an opening of the gap between conduction and valence band which touch each other at the zone edge K point of brillouin zone \cite{CastroNetoReview}. This on-site energy difference can be controlled externally by the application of   electric field, perpendicular to the layers, which implies a potential energy difference $\Delta$V and hence a gap $\Delta_g$. Experimentally such fields in-between two carbon layers were created either by application of  top and bottom gates simultaneously \cite{Morpugo,Atin,NatureIR} or by chemical doping of a back gated device \cite{GeimBiased,OhtaScience}. In the presence of an electric filed in between the layers, the Hamiltonian for the bilayer graphene near the K point can be written as 

\begin{equation}
\label{Hamiltonian}
\mathbf{H}=\bordermatrix{& \cr          
& -\Delta{V}/2  & \gamma \bf{k}  & 0  & 0 \cr          
& \gamma \bf{k}  & -\Delta{V}/2  & \gamma_{1}  & 0 \cr       
& 0  & \gamma_{1}  & \Delta{V}/2  & \gamma \bf{k} \cr
& 0  & 0  & \gamma \bf{k}  & \Delta{V}/2 \cr}
\end{equation}
with, $\gamma=\frac{\sqrt{3}}{2}\gamma_{0}a$ where $\gamma_0$ and $\gamma_1$ are the inplane and interlayer nearest neighbor hopping energies respectively.
The eigen-values of the above Hamiltonian for the lower subbands of the bilayer graphene can be written as :
\begin{equation}
\epsilon_{\pm}(k)=\pm\sqrt{\left(\frac{(\Delta{V})^{2}}{4} + \frac{\gamma_{1}^{2}}{2} + \gamma^{2}{k}^{2}\right) - \sqrt{\frac{\gamma_{1}^{4}}{4} + \gamma^{2}{k}^{2}\left(\gamma_{1}^{2} + (\Delta{V})^{2}\right)}}
\end{equation}
where $\pm$ correspond to conduction and valence bands, respectively. Figure \ref{FigGap}(a) shows the band structure of bilayer graphene having a band gap ($\Delta_g$). The relation between the band gap and potential energy difference ($\Delta$V) is given by \cite{CastroNetoReview}

\begin{equation}
  \label{gap}
  \Delta_g = \sqrt{\gamma_1^2 (\Delta V)^2/(\gamma_1^2+(\Delta V)^2)}
  \end {equation}
which shows that there will be no band gap if $\Delta$V = 0.

Figure \ref{FigGap}(b) shows the induced carrier density in each carbon layer along with the electric field in between them in the presence of top and back gates. The total induced carrier density in bilayer, due to $V_{T}$ is: $n_{T}e = - C_{T} V_{T}$ and similarly, for $V_{B}$, $n_{B}e = - C_{B} V_{B}$. Therefore, total induced carrier density in bilayer graphene due to top and back gates is: 
\begin{equation}
\label{inducedcharge_woscreening}
ne = n_{T}e + n_{B}e = -\left[C_{T}V_{T} + C_{B}V_{B}\right]
\end{equation}
where, for simplicity, we have neglected the quantum capacitance . Considering the screened electric field , the induced carrier density due to top gate is $n_{T_{1}}$ and $n_{T_{2}}$ where subscripts 1 and 2 refer to top and bottom layers, respectively, as shown in figure \ref{FigGap}(b). For a screening length of $\lambda$, we can write for top gate n$_{T_{2}} = n_{T_{1}}e^{-d/\lambda}$ where $d$ is the  separation between the two layers ($3.41 \sf{\AA}$). Thus, for the top gate $n_{T} = n_{T_{1}} + n_{T_{2}}$ and $n_{T_{2}} = \frac{n_{T}} {1+ e^{d/\lambda}}$. Similar expression holds good for the back gate. Thus, in presence of back and top gates the resultant electric field inside the bilayer would be $\vec{E} = \vec{E}_{T} + \vec{E}_{B}$, where 

\begin{eqnarray}
\label{field_screening_top}
&&E_{T} = \frac{n_{T}e}{2\epsilon} - \frac{n_{T_{1}}e}{2\epsilon} + \frac{n_{T_{2}}e}{2\epsilon} \nonumber \\
&&	=  \frac{n_{T_{2}}e}{\epsilon} = \frac{n_{T}e}{\epsilon(1+e^{d/\lambda})} =  \frac{C_{T}V_{T}}{\epsilon(1+e^{d/\lambda})}
\end{eqnarray}
where $\epsilon$ $\sim$ 2.5 is the dielectric constant of the bilayer graphene \cite{CastroNetoReview}.

The same analysis holds good for $E_{B}$ as well. Therefore, the net electric field between two carbon layers is, 

\begin{equation}
\label{field_screening}
\vec{E} =  -\hat{y}\left[ \frac{C_{T}V_{T}}{\epsilon(1+e^{d/\lambda})} -  \frac{C_{B}V_{B}}{\epsilon(1+e^{d/\lambda})}\right]
\end{equation}

It should be noted that  for $\lambda = \infty$ the equation (\ref{field_screening}) reduces to 

\begin{equation}
\label{field_woscreening}
\vec{E} = -\hat{y}\left[ \frac{C_{T}V_{T}}{2\epsilon} -  \frac{C_{B}V_{B}}{2\epsilon}\right]
\end{equation}

i.e. the field without screening which corresponds to two layers having equal carrier density.

It can be seen from equation (\ref{inducedcharge_woscreening}) and (\ref{field_screening}) that minimal value of the resistance maximum should appear when V$_{B}$ = 0 V and V$_{T}$ = 0 V, since this would imply that Fermi energy E$_F$ = 0 (since $n$ = 0) and band gap $\Delta$g = 0 (since E = 0). As mentioned earlier, in our experiment minimum value of R$_{m}$ appears at V$_{T}$ = -0.4 V (figure \ref{Fig1}) because of unintentional electron doping. To consider the effect of electric field inside the bilayer due to this unintentional electron doping, we place a positive charge sheet of carrier density $n_0$ above the top carbon layer, given by $n_ 0 = C_{B}V_{B}^*$ obtained from the back gate sweep for V$_{T}$ = 0 V, as shown in figure \ref{Fig1} (V$_{B}^*$$\sim$ -40V). Thus the net electric field inside the bilayer is written as 

\begin{equation}
\label{field_screening_n0}
\vec{E} = \vec{E}_{T}	+ \vec{E}_{B} + \vec{E}_0 = -\hat{y}\frac{1}{\epsilon(1+e^{d/\lambda})}\left[C_{T}V_{T} -  C_{B}V_{B} +  C_{B}V_{B}^*\right]
\end{equation}
where $\vec{E}_0$ is the field due to $n_0$. The potential energy difference between the two layers is $\Delta V = e(E \times d)$, and the value of the gap is estimated from equation (\ref{gap}). Taking back gate capacitance C$_{B} = 12 nF/cm^{2}$ and top gate capacitance C$_{T} \sim 1.5 \mu F/cm^{2}$, figure \ref{FigGap}c shows the plot of the value of the band gap, with screening $\lambda = 5 \sf{\AA}$ \cite{Rise Nov} and without screening $\lambda = \infty$, as a function of V$_{T}$ (where band gap values are evaluated at R$_{m}$ for each of the back gate characteristic at fixed V$_{T}$ (figure \ref{Fig1})). It can be seen from figure \ref{FigGap}c, that $\Delta_g$ is minimum at V$_{T}$ = - 0.4 V. This trend follows from the trend of R$_{m}$ (see figure \ref{Fig1}). We notice that, had the arrangement of $n_0$ been placed below the bottom layer, then the field components associated with it would have changed their direction and the calculated energy gap would not have followed the trend of R$_{m}$. Thus, our assumption that the arrangement of unintentional carrier density $n_0$ lies above the top layer has indeed been justified

\section{Conclusions}

In conclusion, the bilayer channel transforms from n type to p type conduction channel as we vary the drain-source voltage. This variation results in nonlinear dependence of I$_{DS}$ on V$_{DS}$ as a function of V$_T$ which has been quantitatively explained. Contrary to the recently reported monolayer case, we did not observe any signature of current saturation for I$_{DS}$. In the presence of a gap in bilayer energy spectrum, the formation of p-n junction creates new possibility for the bilayer device as a source of terahertz radiation \cite{TeraMag, TeraJETP}, similar to infrared emission from the carbon nanotubes \cite{CNTScience}. The radiation from recombination between electrons and holes at the junction may acts as terahertz source depending on the magnitude of the gap that can be opened. The opening of a gap in the bilayer spectrum was studied, where the gap was controlled using a top and back gate geometry. However the band gap opening is limited by the restriction on the back gate voltage, since the breakdown field for the SiO$_2$ is 1V/nm ( up to 300V in our case). The capacitance of the polymer electrolyte was determined accurately. Thus solid polymer electrolyte gating of graphene devices may open up new avenues for graphene based electronics, particularly where higher carrier concentration is required.

\begin{acknowledgments}
AKS thanks Department of Science and Technology for financial assistance through the DST Nanoscience Initiative Project.
\end{acknowledgments}

\end{document}